\documentclass[fleqn, usenatbib]{mnras}

\usepackage{newtxtext, newtxmath}
\usepackage[T1]{fontenc}

\usepackage{graphicx}
\usepackage{amsmath}
\usepackage{acronym}
\usepackage[separate-uncertainty = true, multi-part-units=single]{siunitx}
\usepackage{amsfonts}
\usepackage{comment}
\usepackage{subcaption}
\usepackage{xspace}

\newacro{MJD}{Modified\ Julian\ Day}
\newacro{TOA}{time\ of\ arrival}
\newacro{JBO}{Jodrell Bank Observatory}
\newacro{GPR}{Gaussian process regression}

\newcommand{\lsl}{\ensuremath{\langle S \rangle}\xspace}
\newcommand{\wten}{\ensuremath{w_{10}}\xspace}

\title{Observation of discontinuities in the periodic modulation of PSR B1828$-$11}

\author[A. Dias et al.]{
Adriana Dias$^{1}$,\thanks{E-mail: adriana.dias@rhul.ac.uk (AD)}
Gregory Ashton$^{1}$,
Julianna Ostrovska$^{1}$,
David Ian Jones$^{2}$,
Michael Keith$^{3}$
\\
$^{1}$Physics Department, Royal Holloway, University of London, Egham Hill, Egham, TW20 0EX, United Kingdom\\
$^{2}$Mathematical Sciences, University of Southampton, Southampton SO17 1BJ, United Kingdom\\
$^{3}$Department of Physics and Astronomy, Jodrell Bank Centre for Astrophysics, The University of Manchester, Manchester M13 9PL, United Kingdom\\
}

\pubyear{2024}

\begin{document}
\label{firstpage}
\pagerange{\pageref{firstpage}--\pageref{lastpage}}
\maketitle

\begin{abstract}
%It should be a single paragraph not more than 250 words (200 words for Letters).
PSR B1828$-$11 is a radio pulsar that undergoes periodic modulations ($\sim$\SI{500}{days}) of its spin-down rate and beam width, providing a valuable opportunity to understand the rotational dynamics of neutron stars.
The periodic modulations have previously been attributed to planetary companion(s), precession, or magnetospheric effects and have several interesting features: they persist over 10 cycles, there are at least two harmonically related components, and the period is decreasing at a rate of about 5 days per cycle.
PSR B1828$-$11 also experienced a glitch, a sudden increase in its rotation frequency, at \SI{55040.9}{\ac{MJD}}. By studying the interaction of the periodic modulations with the glitch, we seek to find evidence to distinguish explanations of the periodic modulation. 
Using a phenomenological model, we analyse a recently published open data set from Jodrell Bank Observatory, providing the longest and highest resolution measurements of the pulsar's spin-down rate data.
Our phenomenological model consists of step changes in the amplitude, modulation frequency, and phase of the long-term periodic modulation and the usual spin-down glitch behaviour.
We find clear evidence with a (natural-log) Bayes factor of 1486 to support that not only is there a change to these three separate parameters but that the shifts occur before the glitch.
Finally, we also present model-independent evidence which demonstrates visually how and when the  modulation period and amplitude change.
Discontinuities in the modulation period are difficult to explain if a planetary companion sources the periodic modulations, but we conclude with a discussion on the insights into precession and magnetospheric switching.

\end{abstract}


%%%%%%%%%%%%%%%%%%%%%%%%%%%%%%%%%%%%%%%%%%%%%%%%%%

%%%%%%%%%%%%%%%%% BODY OF PAPER %%%%%%%%%%%%%%%%%%

\section{Introduction}
\label{sec:introduction} 

Pulsars provide a unique astrophysical laboratory to probe physics at the extreme.
One avenue to better understand pulsars is through the investigation of pulse timing, which may reveal insights into the properties of the magnetosphere (which emits the observed radiation) or the interior of the neutron star itself.
In this work, we study data on the spin-down rate of PSR B1828$-$11 (i.e., the time derivative of the pulsation frequency), performing a phenomenological model fit to study features in a new high-resolution data set recorded at \ac{JBO}.
This pulsar exhibits several interesting and related phenomena:
the timing properties are periodically modulated with a timescale of $\sim$\SI{500}{days} and display a characteristic double-harmonic-sinusoid structure.
Meanwhile, the pulse shape rapidly switches between two distinct states, and the proportion of time spent in each state is also modulated and correlated with the timing variations.
Finally, the modulation period decreases with time, and the star has undergone a glitch - a sudden spin-up event.
This rich mixture of observations requires a unified explanation.
Three primary model interpretations have been proposed: the presence of a planet or system of planets orbiting the pulsar, free precession and magnetospheric switching.
The ultimate goal of this work is to utilise the new data to constrain these models.

\section{Previous studies of PSR B1828$-$11}
\label{sec:history}

\citet{bailes1993planets} reported the first observations of PSR B1828$-$11 and hypothesised a planetary explanation, noting that a system of at least two planets would be required to explain the two harmonics observed in the timing properties of the star.
However, in \citet{Stairs2000}, an extended data set was analysed, covering several cycles and simultaneously analysing timing properties and the pulse shape (via an averaged shape parameter \lsl), finding strong correlations between the two. 
Based on this observation, \citet{Stairs2000} rejected the planetary explanation since it would require the planet, orbiting at about 1~AU, to interact with the magnetosphere that is at most a few thousand kilometres.
Nevertheless, recent work by \citet{Liu2007} studied a quark planetary model and separately \citet{Nitu2022} conducted a search for planetary companions around 800 pulsars, finding that PSR B1828$-$11 could, in principle, be explained with two planetary companions (though they conclude that intrinsic spin variation is a better-supported explanation).

Instead, \citet{Stairs2000} proposed free precession as the cause of the periodic modulation.
They postulated that the periodicity of PSR B1828$-$11, with harmonically related sinusoids with periods of $\sim$ \SI{1000}{}, \SI{500}{} and \SI{250}{days}, was resultant from precession of the spin axis, caused by the misalignment of the angular momentum and symmetry axis of the star and assuming the star to be non-spherical.
This work was followed by physical models proposed by \citet{Jones2001} and \citet{Link_2001}, where the authors explored how the variations in the pulse shape and timing of PSR B1828$-$11 could be explained by free precession of the star's crust causing variations in the magnetic dipole torque angle.
They found the observations could be explained by a star precessing with a period of $\sim$ \SI{500}{days} and a wobble angle of $\sim$ \SI{3}{\degree}, assuming that the magnetic dipole is nearly orthogonal to the star's symmetry axis.
This configuration is somewhat special as it means that the dipole cuts through the equator four times per precession period, producing the characteristic double-harmonic-sinusoid observations (see Fig.~\ref{fig:data_old_new_superimposed}).
Moreover, \citet{Link_2001} fitted the model to the data and found that an hourglass-type beam geometry was required to explain the observed \lsl data.
Further advances of precession include a tri-axial body with core and blob beam geometry \citep{Akgun2006} and the development of a time-varying magnetic field \citep{Rezania2003}.

Following further observations, the free precession interpretation was challenged in \citet{Lyne2010}. Most notably, they highlight that the time-averaging baseline required to measure the spin-down rate (used in the beam-shape parameter of \citet{Stairs2000}) will obscure behaviour happening on faster timescales.
Following the contemporaneous identification of rapid magnetospheric switching phenomena (see, e.g. the extreme case of PSR B1931+24 \citet{Kramer2006}, where the pulsar switches on and off with correlated changes in its spin-down rate), the authors proposed that the spin-down and beam-width variations of PSR B1828$-$11 could similarly be explained by a model in which the magnetosphere switches between states in a quasi-periodic fashion, but that the probability of being in one state or the other varies on the modulation timescale.
This suggestion is based on the identification \citep{Stairs2000, Stairs:2002vz} that the pulsar exhibits distinct narrower and wider profiles.
They explained that inferred parameters such as the spin-down rate and shape parameter \lsl, which use a multiday baseline, average over behaviour on shorter timescales, revealing the slow time-varying probability between states but obscuring the rapid switching.
To evidence this, they introduce a new pulse shape parameter, \wten, that could be calculated on individual observations to avoid the longer 100-day baseline required to measure the spin-down.
A further follow-up study of PSR~B1828$-$11 in \citet{Stairs2019} used additional high-solution observations from the Parkes and Green Bank Telescopes, which enabled a detailed study of the pulse-to-pulse behaviour.
They confirmed that there are only two distinct pulse shapes. By correlating the ratio of the time spent in each state with the modulation phase, they further validate the model proposed by \citet{Lyne2010}.

A generative model of the switching process was developed in \citet{Perera2015} and applied to the PSR B0919+06, which shows a similar pattern of long-term behaviour to PSR B1828$-$11.
To explain the characteristic double-harmonic-sinusoid present in the spin-down rate of B0919+06 with a two-state magnetospheric switching model, \citet{Perera2015} proposed a \emph{four-phase model} in which the pulsar switches between the two states twice per cycle. The characteristic second (lower) peak arises because the time spent in the state is shorter than the time-averaging window used to generate the spin-down data. The authors of \citet{Shaw2022} also report a similar behaviour in PSR B0740-28, whose profile exhibits two distinct shapes. 
Taking this model, \citet{Ashton2016} compared the precession and switching hypotheses for PSR B1828$-$11, analysing the spin-down and \wten pulse shape data from \citet{Lyne2010}. They augmented the standard precession model with a variable braking index and included a flexible beam profile.
Meanwhile, the four-state switching model proposed by \citet{Perera2015} was applied, additionally modelling the time-averaging process to predict the spin-down and connecting each state with a separate beam width \wten.
\citet{Ashton2016} concluded that, based on the models and data under consideration, precession was the favoured explanation.

However, the more recent study of \citet{Stairs2019} points out that the precession beam-width model applied in \citet{Ashton2016} is at odds with the observation. Specifically, \wten varies slowly due to changes in the line-of-sight view of the radio emission, while the data demonstrates that it, in fact, varies rapidly between pulses.
Nevertheless, while the pulse shape model of \citet{Ashton2016} is mistaken, the precessional explanation of the spin-down is the more parsimonious: it both provides a natural clock and avoids the complicated four-phase model required to explain the double-harmonic-sinusoid.
Moreover, as pointed out by \citet{Jones2012}, it remains plausible that precession is the clock driving the long-timescale variability. For example, the spin-down could remain a product of the effects of precession, while the time-varying wobble of the star could be responsible for driving the unstable magnetosphere to switch between quasi-stable states.

We also point out in reviewing \citet{Ashton2016} that the \citet{Perera2015} four-phase switching model also seems to be at odds with our new observations. 
Specifically, in this model, the pulsar switches rapidly between two distinct spin-down rates but switches twice per cycle.
To produce the double-peaked spin-down rate, the time spent in one of the states must be shorter than the time-averaging baseline.
It therefore follows that the secondary peak ``height'' is a function of the time-averaging baseline. If the baseline is sufficiently short, there will be a double peak, but the heights will be equal; only the duration spent in each state will differ.

In Fig.~\ref{fig:data_old_new_superimposed}, we can compare the data produced using the \SI{100}{day} baseline from \citet{Lyne2010} with the higher-resolution data obtained generated by \citet{Keith2023} using a Fourier-basis \ac{GPR} (described later in Section~\ref{sec:data}).
Notably, we do not see a variation in the height of the second peak.
While the methods are not directly comparable, the average time between TOAs in the data analysed by \citet{Shaw2022} was \SI{7}{days}; since in \citet{Ashton2016}, it was shown that the duration spent in the short-duration state was approximately 10-20~days, we would therefore expect the \citet{Keith2023} inferred spin-down measurements to be more sensitive to the step-changes if the pulsar switches suddenly and semi-permanently between states as in the \citet{Perera2015} model.
However, this is not the case.
Therefore, this observation is inconsistent with the four-phase switching model and suggests that whatever mechanism drives the spin-down variations smoothly varies between the minima and maxima (as previously argued and demonstrated by \citet{Stairs2019}).
To account for this observation, the four-phase switching model could be modified. 
Minimally, one could introduce three distinct states, though this then causes inconsistencies with the observation of the beam width, which itself does not show any evidence of a third state.

The complexity of B1828$-$11 became more interesting when \citet{Ashton2017} discovered that the modulation period present in the spin-down data is itself getting shorter, losing about 1 day per 100 days, and identified from the Jodrell bank glitch catalogue \citep{Basu2021} that the pulsar also experienced a glitch, a sudden increase in the rotation frequency, at \SI{55040.9}{\ac{MJD}} (coinciding with the end of the data set provided by \citet{Lyne2010}).
Leading models of glitches suggest they provide evidence for a superfluid component in the core of the star.
However, such a superfluid component is incompatible with precession \citep{Shaham1977, Jones2001, Haskell:2023exo} since the pinning of the superfluid would result in a free precession period much shorter than the observed modulation period of $\sim 500$ days, and may also be expected to be rapidly damped.

The implications of this were discussed in \citet{Jones2017}, where several models tried to tie together the decreasing modulation period with the glitch, making predictions for the subsequent behaviour. In the main, these predicted that the glitch should produce changes in the modulation period.
However, subsequent analyses \citet{Brook2016, Stairs2019, Shaw2022} have demonstrated that the modulation of the timing properties on a $\sim 500$~day timescale continues after the glitch.
However, to date, no quantitative study has been performed to determine if there are any step changes associated with the glitch.

Very recently, \citet{lower_etal_25} made a study of radio emission variability in a sample of $259$ pulsars, making two findings of potential relevance here.  Firstly, they found that variations in both spin-down rate and pulsar profile shape are more common than previously thought.  Secondly, by looking at the set of $45$ pulsars that exhibit quasi-periodic variations in their spin-down date, they found that the modulation period of the variations was approximately independent of the spin-period, a result not expected on the basis of several free precession models described in \citet{Jones2012}.  This last point makes the free precession interpretation of quasi-periodic timing variability less attractive, at least as a common explanation for all such variable pulsars.

In any case, it remains unclear what mechanism is responsible for the long-term behaviour of PSR B1828$-$11. While we can agree that the magnetosphere switches rapidly between two states and that this varies coherently on a 500-hundred-day cycle with variations in the spin-down, we do not yet know ``what sets the clock of this cycle?'' ``If it is switching between just two states, why is the spin-down smoothly varying?'' and ``Why is the modulation period decreasing, and are there any changes related to the glitch?''

To answer these questions, we revisit the analysis of PSR B1828$-$11 using the new high-resolution spin-down data \citep{Keith2023}. And, to avoid pre-conditioning our interpretation with a physical model, we apply a phenomenological model to capture the salient features that may be present in the spin-down rate of the pulsar. We will model the spin-down rate data for this pulsar and ascertain whether any step changes occur around the glitch that changes its spin-down rate or modulation. To consider several possibilities, we developed three models to describe the spin-down rate of this pulsar: a model which assumes that a glitch occurred and that there are changes to the periodic modulations; a model which assumes that there is no glitch nor changes to the periodic modulation and another one which assumes there is a glitch but no changes to periodic modulation. We obtained and compared the natural-log evidence for these three models to understand which one fits the data more appropriately. 

The paper is structured as follows.
We first introduce the data set and methodology in Sections~\ref{sec:data} and \ref{sec:data_analysis_model} before describing the models and the fits to the data in Section~ \ref{sec:models}. Then, in Section~\ref{sec:lomb_scargle}, we study the time-period behaviour of the pulsar and compare this with the features extracted from the model. Finally, we conclude with a discussion and outlook in Section~\ref{sec:discussion} and \ref{sec:outlook}, respectively.

\section{Data}
\label{sec:data} 
In this work, we will analyse the open spin-down data published in \citet{Keith2023} (and available from \citet{keith_2023_7664166}), which was derived using a Fourier-basis \ac{GPR} on the raw data in \citet{Shaw2022}.
Observations were conducted using the \SI{76}{\meter} Lovell telescope and were supplemented with data from the \SI{25}{\meter} ``Mark-II'' telescope, both located at \ac{JBO} \citep{LOVELL1957}. Data collected before 2009 was centred at \SI{1400}{\mega \hertz} and recorded using a \SI{32}{\mega \hertz} filterbank. After 2009, data collection shifted to being centred at \SI{1520}{\mega \hertz} and recorded with a \SI{384}{\mega\hertz} filterbank. Detailed information on data acquisition settings can be found in \citet{Shaw2022}.
To transform the acquired data into the spin-down rate analysed in this paper, \citet{Shaw2022} generated a single pulse profile for each observation epoch by summing the data across all frequency channels. The \ac{TOA} is obtained by comparing this integrated pulse profile with a high signal-to-noise profile representing the observed profile's expected shape.
The \ac{TOA}s are then fitted with a timing model \citep{Hobbs_2006}, then subtracting this model from the data results in a timing residual.
Finally, the timing residual is fitted using Fourier-basis \ac{GPR} and from this, the second derivative of the spin-down $\dot{\nu}$ is extracted (see \citet{Keith2023} for further details).

\begin{figure*}
	\includegraphics[width=\linewidth]{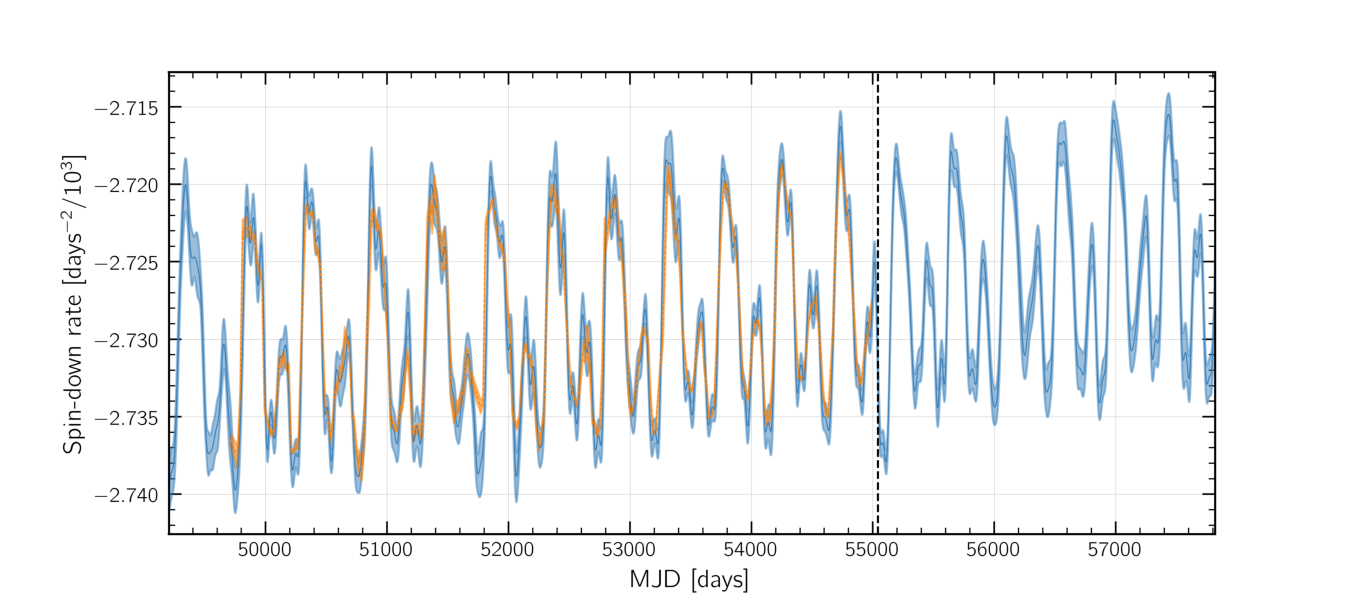}
    \caption{Comparison between \protect\cite{Lyne2010} (in orange) and \protect\citet{Keith2023} (in blue, used in this work) datasets of spin-down rate, with respect to time in MJD. The black dashed vertical line highlights the glitch time for PSR B1828$-$11.}
    \label{fig:data_old_new_superimposed}
\end{figure*}

Previous studies on PSR B1828$-$11, i.e. in \citet{Ashton2017}, utilised a smaller dataset spanning \SI{5280}{days} 
between \SI{49710}{\ac{MJD}} and  \SI{54980}{\ac{MJD}}, which ended before the glitch occurred.
This dataset had 755 \ac{TOA}s, and the spin-down rate was obtained by applying a timing model to a sliding window of duration \SI{100}{days} over the data \citep{Lyne2010}.
In contrast, the \citet{Keith2023} dataset used in this paper spans \SI{8615}{days} between \SI{49202}{\ac{MJD}} and \SI{57817}{\ac{MJD}}, encompasses the glitch event, and has better resolution.
Fig.~\ref{fig:data_old_new_superimposed} illustrates the differences between these datasets, with the old dataset shown in orange, the newer dataset in blue and the glitch time marked by a vertical black dotted line.

\section{Data Analysis Methodology}
\label{sec:data_analysis_model} 

This section provides a brief overview of the Bayesian methodology we use to analyse the data under a set of phenomenological models.
(For a general introduction, see, e.g. \citet{edition2013bayesian}.)

Bayes theorem aims to solve the inverse problem: what can be learnt about model $M$ and its associated parameters \textbf{$\vartheta$}, based on data \text{d}? This can be described by Equation \ref{eq:1}: 
\begin{equation}
    p(\vartheta | \text{d}, M) = \frac{\mathcal{L}(\text{d}|\vartheta, M)\pi(\vartheta|M)}{\mathcal{Z}(\text{d}|M)}\,,
    \label{eq:1}
\end{equation}
where $ p(\vartheta | \text{d}, M)$ is the posterior probability distribution of the parameters $\vartheta$ given the data and the model; $\mathcal{L}(\text{d}|\vartheta, M)$ is the likelihood function of the data, given the parameters and the model; $\pi(\vartheta|M)$ is the prior probability distribution, associated with the set of model parameters; and $\mathcal{Z}(\text{d}|M)$ is the evidence for the data, given the model, and can be calculated from $\mathcal{Z}(\text{d}|M) = \int_{\vartheta} \mathcal{L}(d|\vartheta, M)\pi(\vartheta|M)d\vartheta$. 

We will use the Bilby Bayesian inference library \citep{Ashton2019} to draw samples from the posterior probability density and estimate the evidence using the nested sampling algorithm \citep{Skilling2004}, specifically, the \texttt{dynesty} sampler \citep{Speagle2020,sergey_koposov_2024_12537467}.
Nested sampling enables efficient exploration of the multi-modal and higher-dimensional space we will explore, producing a set of samples approximating the posterior $p(\vartheta | \text{d}, M)$ and an estimate of the evidence $\mathcal{Z}(d| M)$ which we will use for model comparisons.

In contrast to previous works that used Bayesian analyses, throughout this work, we will use ‘slab-and-spike’ priors \citep{Malsiner-Walli_Wagner_2016}.
These comprise a \textit{slab}, usually a standard prior distribution, such as a uniform or a normal distribution prior, and a Dirac \textit{spike} at a fixed location.
We use these in our phenomenological model as a means to marginalise over the model dimensionality without requiring the implementation of a transdimensional sampler \citep{green2003trans}. 
Consider a polynomial of degree $N$ with coefficients $a_i$ with $ i \in [0, N]$ as a generic example. A naive analysis may apply a Bayesian analysis to each degree, treating each as a separate ``model''; a transdimensional sampler improves on this by including $N$ as a model parameter, enabling automatic marginalisation over the model size. However, implementation is often domain-specific (though see \citet{Tong2024}).
Instead, slab-spike priors can be used with regular stochastic samplers when the models are nested (e.g. in the polynomial case, a model of degree $N$ is equivalent to a model with degree $N-1$ with the parameter $a_N$ fixed to zero).
By placing the spike at the point that recovers the simpler model (e.g. $a_i=0$), higher-dimensional models can be explored with the sampler finding posteriors equal to zero for higher-dimensional parameters that don't improve the fit.

\section{Defining and fitting models}
\label{sec:models} 

In this section, we define three phenomenological models of the secular spin-down and periodic modulations to fit the data in Fig.~\ref{fig:data_old_new_superimposed}.
For all three models, the secular part encodes a standard expansion of $N_s$ frequency derivatives, and the periodic modulations utilise a sinusoid with $N_c$ harmonically related components.
Within each sinusoidal term, the phase follows an expansion up to the $N_f$ phase-derivate to capture the slow changes to the modulation period observed in \citet{Ashton2017}.
We start with the most general model, referred to as \textbf{Model: S + P}, which allows independent step changes in the secular spin-down and periodic modulation. 
We also explore two subsets of the \textbf{S+P} model: one which assumes that there is no glitch nor changes to the secular spin-down or the periodic modulation (\textbf{Model: no-glitch}) and another one which assumes a step change only in the secular spin-down  (\textbf{Model: S}).
These subset models allow us to probe the significance of changes in the periodic modulation relative to the other step changes.
For each model, we discuss the theoretical reasoning first, then explain the choice of priors and, finally, the inferred posteriors. 

\subsection{\textbf{Model: S+P}}
\label{sec:glitch_model} 

In this section, we define and apply a model in which a glitch occurs (modelled by an instantaneous change in the spin-down rate accompanied by a transient decay) and that there are also instantaneous changes to the features of the periodic modulation.
We model changes to the features of the periodic modulations as step functions and allow a step change in each component separately.

To develop a full generative model, first we define $t'=t - t_0$, where $t$ is the \ac{MJD} of the observed data and $t_0$ is the \ac{MJD} of a reference time (\SI{55372}{\ac{MJD}} as quoted in \citet{Parthasarathy2019}).
We then write the spin-down rate as:
\begin{align}
\dot{\nu}(t) = &  
\sum_{i=0}^{N_s-1} \frac{\dot{\nu}_i}{i!}\left[1 + H\left(t' - t_s^\xi\right)\left(\xi^p_i + \xi^t_i e^{-\frac{t' - t_s^\xi}{\tau_i}}\right)\right] \Delta t^{i} \nonumber \\
& + \sum_{j=1}^{N_c}A_j\left[1 + H(t' - t_s^\eta)\eta_j\right] \nonumber \\
& \cdot \cos\left(j \phi\left(t\right) + \Delta \phi_j\left(1 + H(t' - t_s^\delta) \delta_j\right)\right)
\label{eq:model_s_p}\,,
\end{align}
where the phase is given by
\begin{equation}
   \phi(t) = 2\pi \sum_{k=0}^{N_f-1} \frac{1}{k !} f_k \left(1 + H(t' - t_s^\chi)\chi_k\right) \Delta t^{k + 1} 
   \label{eq:phase_glitch}\,.
\end{equation}

The key components of this model are: $\dot{\nu}_i$, the $i$th coefficient of the spin-down expansion, $\dot{\nu}_i = \frac{d^{(i)}\dot{\nu}}{dt^{(i)}}$; $A_j$, the $j$th cosine component coefficient (amplitude); $\Delta \phi_j$, the phase-offset of the $j$th cosine component and $f_k$, the $k$th derivative of the modulation frequency.

Within this model, the parameters $N_s$, $N_c$ and $N_f$ define the maximum number of components included in the model.
Ideally, we would like to marginalise over these parameters (e.g. using a trans-dimensional sampler). However, in practice, we will use a maximum value and then apply slab-spike priors.
To determine the maximum value, we analysed the data using the \textbf{S+P} model, incrementing each maximum until no improvement in the fit was found (as quantified by the change in the Bayesian natural-log evidence).
Using this approach, we selected maximum values of $N_s = 3$, $N_c = 8$ and $N_f = 2$; the choice of upper-limit on the number of frequency components is consistent with the frequency range of the Fourier-basis \ac{GPR} used to generate the data \citep{Keith2023}.
To confirm these were sufficiently large, we then verified that the amplitude parameter of the largest component had the maximum posterior support at zero (see Table~\ref{tab:priors_glitch_summary}); in other words, the model preferred a simpler model, and our results are robust to increases in the maximum values.

To model the step changes in each component of Equations~\ref{eq:model_s_p} and \ref{eq:phase_glitch}, we utilise a Heaviside step function multiplying a dimensionless relative amplitude for the spin-down $\xi^p$, the amplitude $\eta$, modulation phase $\delta_j$ and modulation frequency $\chi_k$. There is one step-change per summation term, but all parameters in a given set have a single associated time (i.e. $t_s^\xi$, $t_s^{\eta}$, $t_s^{\delta}$, and $t_s^{\chi}$) such that, e.g. all spin-down changes happen at the same time.
For the changes to the secular spin-down rate, following standard approaches to modelling glitches (see, e.g. \citet{lorimer2005handbook}), there is a permanent offset $\xi^p$ and an exponentially-decaying transient term $\xi^t$ with associated time-scale $\tau$. However, the transient component is only non-zero for the zeroth-order term as set by the priors discussed shortly.

The modulation component of the model follows a harmonic-sinusoid form with an amplitude $A_j$, harmonic phase $j \phi(t)$ (where $\phi(t)$ is the standard phase expansion), and phase offset $\Delta\phi_j$. The harmonic coefficient $j$ multiplies the phase in the argument of the cosine but does not multiply the phase offset. This prevents degeneracies in the solution as $\Delta\phi \in [0, 2\pi]$ while still exploring the entire parameter space. 

\paragraph*{Priors}
We list the complete set of priors used in Table~\ref{tab:priors_glitch_summary}.
For many parameters, we use a uniform prior, choosing a suitable range to cover the expected behaviour (and check where performed to ensure the range did not arbitrarily limit the model fit). We then augment several of these with slab-spike priors emulating a transdimensional sampler.

For the glitch time parameter $t_s^\xi$ affecting the secular spin-down, we apply a prior width ranging $\pm$ \SI{50}{days} around the \SI{55040.9}{\ac{MJD}} based on the recorded glitch time \citep{Basu2021}.
Meanwhile, for the other step-change time parameters, we sample in an offset time relative to $t_s^{\xi}$: that is we define $\Delta t_s^{\alpha} = t_s^{\alpha} - t_s^{\xi}$ for $\alpha \in \{\chi, \delta, \eta\}$ and then apply a uniform prior on $\Delta t_s^\chi$, $\Delta t_s^\delta$ and $\Delta t_s^\eta$ from $-5000$ to \SI{2000}{days}.

For $\dot{\nu}_0$, we apply a wide prior ranging from the minimum to the maximum values of the observed spin-down data shown in Fig.~\ref{fig:data_old_new_superimposed}, i.e. from $-2.74\times10^{-3}$ to $-2.72\times10^{-3}$\SI{}{days ^{-2}}. For all higher-order derivatives of $\dot{\nu}_0$, we set a uniform prior on an arbitrary range and verify the choice of the prior range has no impact on the analysis.

For $\xi_p^0$ and $\xi_t^0$, we set a uniform prior with a range $\pm 0.01$ and again verify this arbitrary range is sufficiently broad.
For $\tau$, we apply a uniform prior between 0 and \SI{500}{days}, ensuring the relaxation time is positive while choosing an arbitrarily large upper value.

The amplitude terms, $A_j$, are given prior distributions ranging from 0 to $10^{-5}$; while a negative amplitude is, in principle, physical, this would introduce degeneracy with the phase term.
The modulation phase offset terms are given a uniform prior on $-\pi$ to $\pi$.
For the step-change parameters, we apply a uniform prior from $-1$ to $1$ for $\chi_j$; we set a uniform prior on $-1$ to $1$; this enables direct interpretation of the posterior without concern about the effects of the prior.
However, for $\eta_j$ and $\delta_j$, we found that with a uniform prior, the sampler failed to robustly identify the maximum-posterior mode (occasionally getting stuck in islands with lower posterior support with larger relative changes. Therefore, we instead apply a standard normal prior such that the prior maximum is zero while setting a scale for expected instantaneous changes, which suppresses order-of-magnitude increases in the amplitude and phase term.

The phase, as seen in Equation \ref{eq:phase_glitch}, contains \textit{k}th derivatives of the modulation frequency. The base modulation period is estimated to be $\sim$ \SI{460}{days}, although as it is shown in Fig.~\ref{fig:lomb_scargle} this modulation period varies from \SI{489}{days} to \SI{435}{days} throughout the entire data range. Thus, we set the prior range of $f_0$ to a range which includes the base modulation frequency, i.e. $ \frac{1}{460}\SI{}{\hertz}$. The other $f_k$ terms have an arbitrary factor of $10^{4+k}$ applied to the modulation frequency.

\begin{figure*}
    \centering
    \subfloat[]{
    \includegraphics[width=\linewidth]{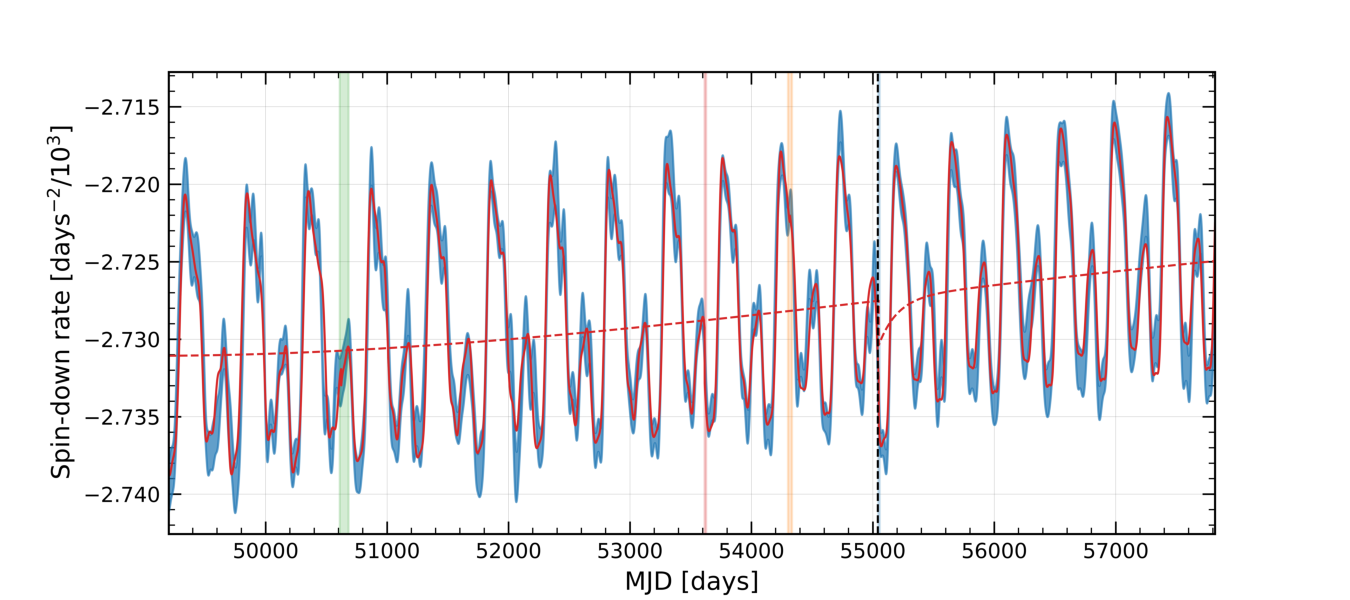}
    \label{fig:data_fit_glitch}} \\
    \subfloat[]{
    \includegraphics[width=\linewidth]{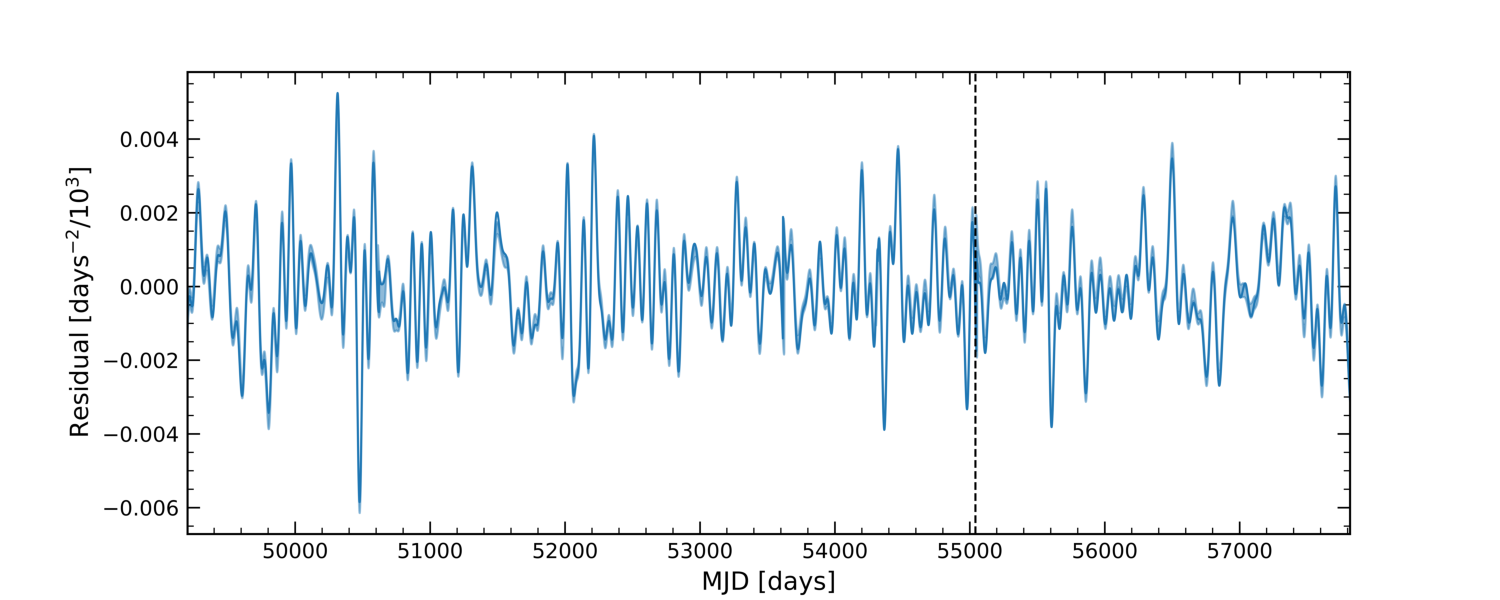}
    \label{fig:residuals_glitch}}
    \caption{\label{fig:data_fit_residuals_glitch}   \ref{fig:data_fit_glitch} shows the spin-down rate data, in blue, together with the maximum posterior estimate solution of \textbf{Model S+P}, in red, which uses the parameters with the highest posterior probability. An orange dotted line shows the spin-down rate component of the model without the modulation cosine components. The glitch time is represented by a black dashed vertical line.  Four vertical shaded 99\% quantile regions are shown, which relate to each of the $t_s$ step-change parameters, with $t_s^\xi$ (the step in spin-down rate) shown in blue, $t_s^\eta$ (the step in modulation amplitude) shown in yellow, $t_s^\delta$ (the step in modulation phase)  shown in green and $t_s^\chi$ (the step in modulation frequency) shown in red. \ref{fig:residuals_glitch} shows the residuals, as a line in blue, obtained by subtracting \textbf{Model S+P} from the data. The blue shaded area around the data shows the 90\% quantile region. Here, a black dashed vertical line also indicates the glitch time. }
\end{figure*}

\paragraph*{Results}
We summarise the posterior distributions in Table~\ref{tab:posterior_glitch_summary}, which contains the median $\pm$ standard deviation values. 
Fig.~\ref{fig:data_fit_glitch} presents the spin-down rate data (in blue) together with the maximum posterior estimate solution of the model (in red) and an orange dashed line showing the secular component of the model alone (i.e. without the periodic modulation) from which we see the analysis identified an exponential recovery present after the glitch.
Additionally, four vertically shaded 99\% quantile regions are shown, which relate to each of the step-change time parameters, with $t_s^\xi$ shown in blue, $t_s^\eta$ shown in yellow, $t_s^\delta$ shown in green and $t_s^\chi$ shown in red. A detailed description of the results of the glitch step change is presented in Section~\ref{sec:non_zero_parameters}. 

In Fig.~\ref{fig:residuals_glitch}, we visualise the residuals obtained by subtracting the model from the data alongside the 90\% interval generated by sampling model draws from the posterior distribution before subtracting for the residual.
We note that, while the broad fit to the data is good, the residual still displays some structure, suggesting further improvements to our phenomenological model are possible.

\subsection{Interpreting the step-changes in inferred parameter for \textbf{Model S+P}}
\label{sec:non_zero_parameters} 

\begin{figure}
    \includegraphics[width=\columnwidth]{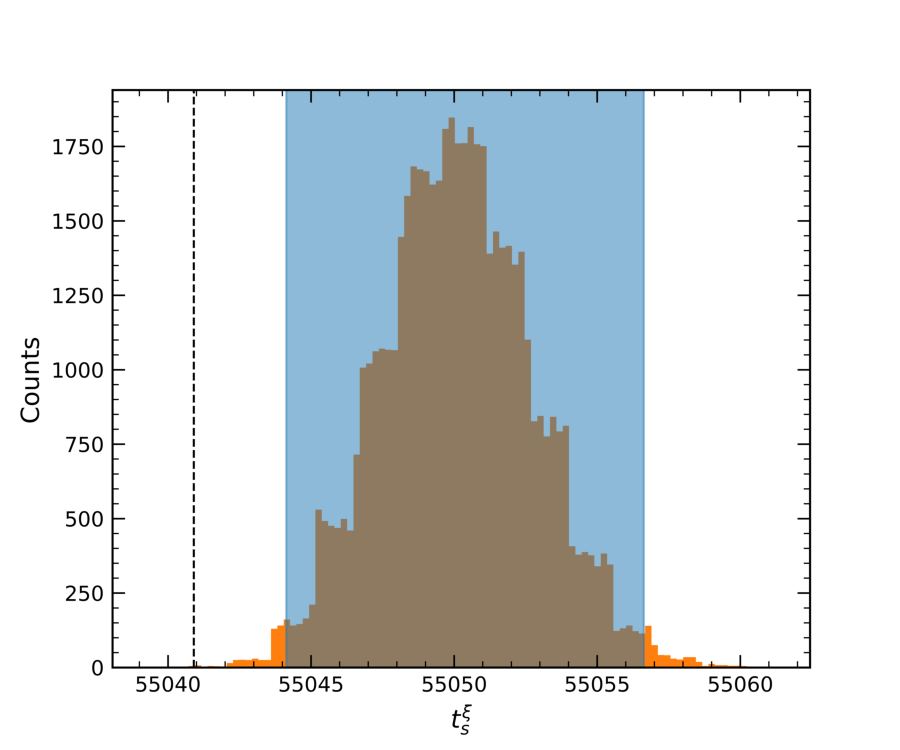}
    \caption{Posterior distribution, shown in orange, for the $t_s^\xi$ parameter that gives the time of the glitch as found in our \textbf{Model S+P}. The 99\% quantile region is shown in blue and the previously reported glitch time of PSR B1828$-$11 is represented by the vertical dashed black line.}
    \label{fig:tg_xi_posterior}
\end{figure}

We find that the posterior distribution of $t_s^\xi$ has a posterior width of $\sim 6$~days at the 99\% credible interval (see Fig.~\ref{fig:tg_xi_posterior}), with a maximum posterior value of 55049, $\sim 9$~days apart from the recorded glitch time of PSR B1828$-$11 (shown as a vertical dashed line in Fig.~\ref{fig:data_fit_glitch}). This difference likely arises from the fact that we are estimating the glitch time from the spin-down rate whereas the glitch time is estimated from the full phase evolution.

For the secular spin-down, we measure the spin-down and its first two derivatives with values consistent with those known in the literature. We also measure a third-order derivative that while non-zero, contained zero at 3 standard deviations.  Fig.~\ref{fig:data_fit_glitch} shows that the model has recovered the step-change and transient recovery observed on the data (which has an inferred timescale of \SI{121(14)}{days)}. We do not allow for step changes in the spin-down derivatives and there is no evidence from the residuals to suggest these are required.

For the periodic modulations, we identify 8 non-zero harmonic components, a significant change relative to the 2 harmonic components that have been fitted to the data before (see, e.g. \citet{Stairs2019, Ashton2016}).
The impact of these higher-order terms can be observed directly in Fig.~\ref{fig:data_old_new_superimposed}: looking at the trailing edge after each of the successive maxima, we can identify in the data a short plateau; this was present in the original data set \citep{Lyne2010}, but is distinct in the newer higher-resolution data analysed in this work.
In Fig.~\ref{fig:data_fit_glitch}, we see the corresponding behaviour of the higher-order terms in the harmonic expansion fitting this feature (this is also present in fits of the no-glitch model as well, c.f. Fig.~\ref{fig:data_fit_no_glitch}).

\begin{figure}
    \includegraphics[width=\columnwidth]{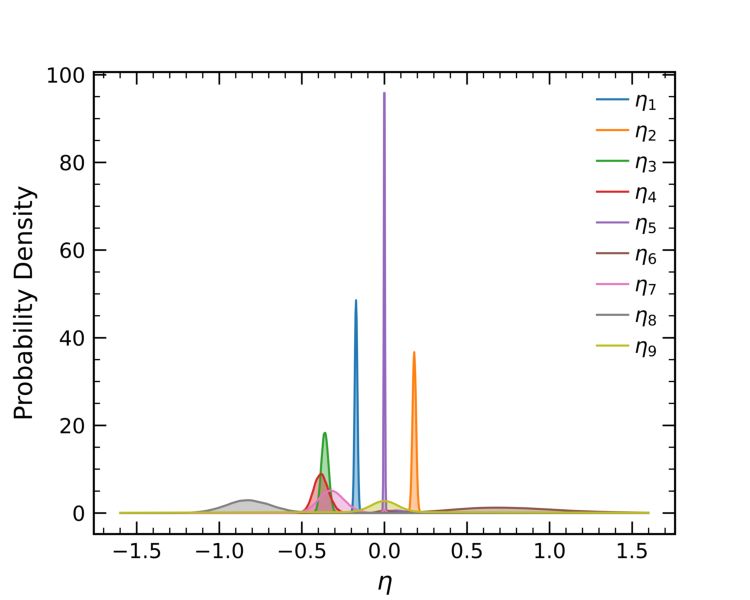}
    \caption{Posterior probability distribution for the $\eta_j$ parameters, for \textbf{Model S+P}.}
    \label{fig:eta_marginals}
\end{figure}

From our analyses, we also identify that a shift in the modulation amplitude occurs at $t_s^\eta = $ \SI{54316.34}{\ac{MJD}} before the glitch occurs.
To visualise the posterior distributions, in Fig.~\ref{fig:eta_marginals}, we plot the posterior distribution for the relative amplitude changes $\eta_{[1-9]}$.
Notably, the first component undergoes a $\sim 20\%$ decrease in amplitude while the second component increases by about the same amount. These two components are the leading order, and the impact can be seen by comparing the fit before and after the glitch in Fig.~\ref{fig:data_fit_glitch}.

We also find evidence for a distinct step-change in the modulation phase at $t_s^\delta= $ \SI{50622.26}{\ac{MJD}}. However, by eye, it is difficult to distinguish in Fig.~\ref{fig:data_fit_glitch} what feature this is fitting: there is no clear discontinuity in the phase at this time.

\begin{figure*}
    \centering
    \includegraphics[width=0.7\textwidth]{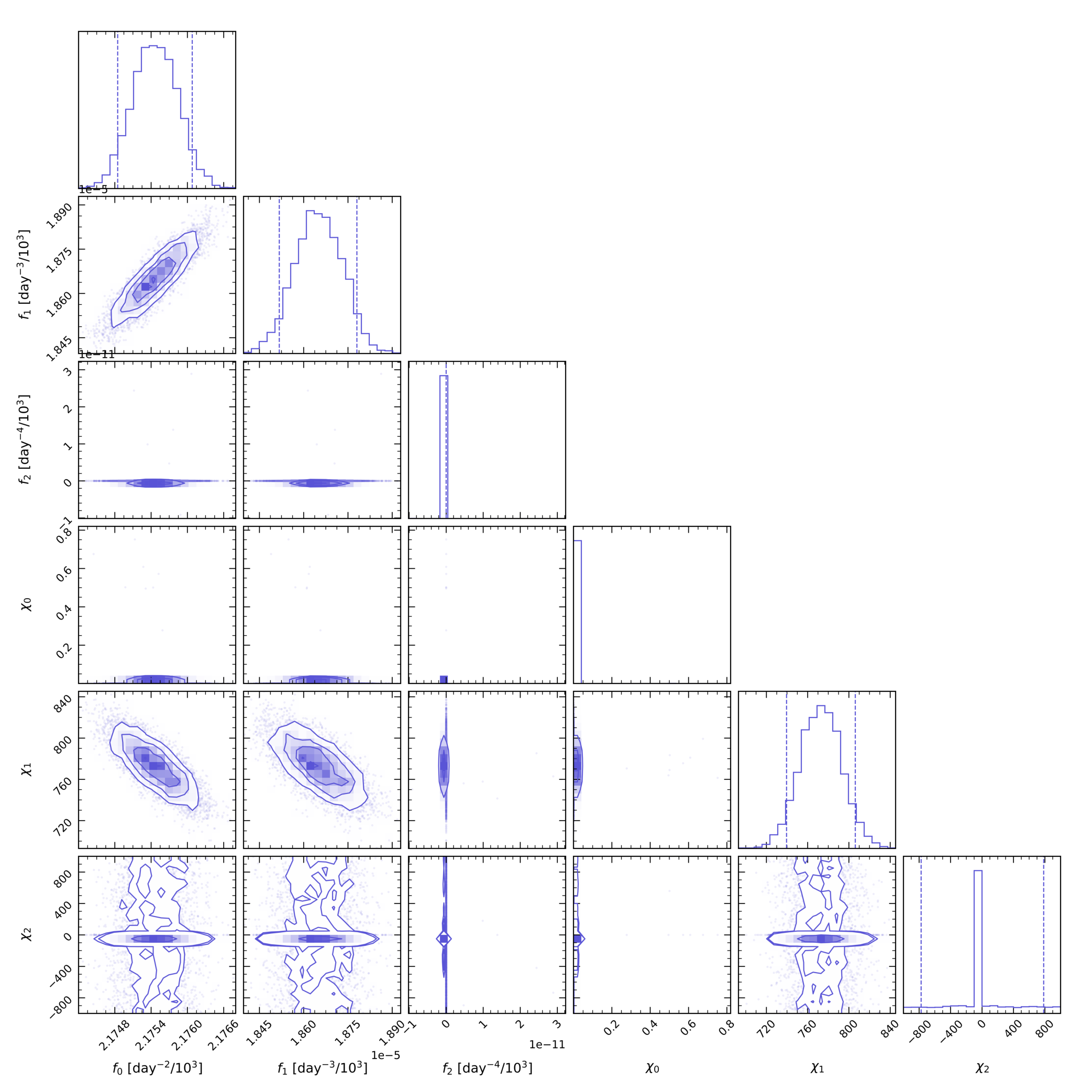}
    \caption{Posterior probability distribution for the modulation frequency terms, $f_k$, and $\chi_k$, which represent their step change, for \textbf{Model S+P}.}
    \label{fig:corner_plot_frequency}
\end{figure*}

\begin{figure}
    \includegraphics[width=\columnwidth]{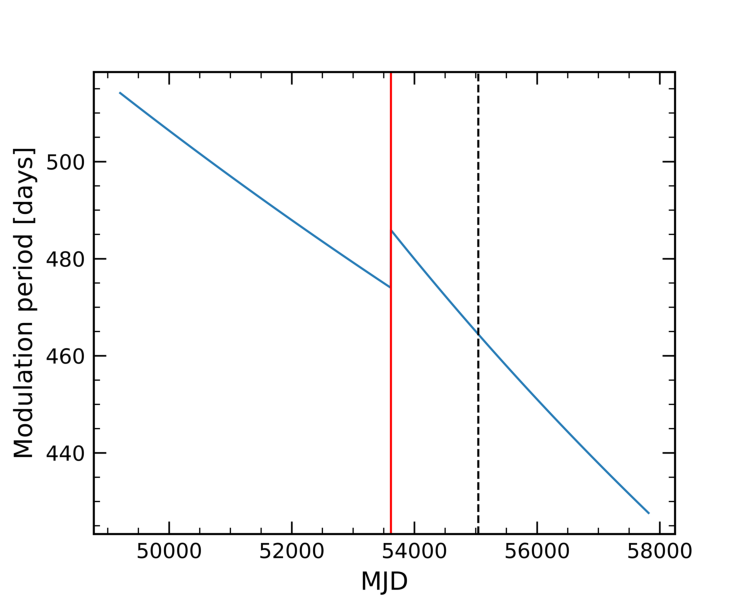}
    \caption{Modulation period vs \ac{MJD}, for \textbf{Model S+P}. The red vertical line indicates the $t_s^\chi$ glitch time parameter, and the black dotted line indicates the glitch time.}
    \label{fig:modulation_frequency_glitch}
\end{figure}

We recover a modulation frequency and non-zero first derivative consistent with values already reported in the literature \citep{Ashton2017}. Our model is also sensitive to a second derivative not previously explored -- Fig.~\ref{fig:corner_plot_frequency}. However, the posterior distribution is consistent with zero, i.e. we do not find any evidence for a second derivative of the modulation period.
We find evidence for a distinct step change in the modulation frequency at $t_s^\chi =$ \SI{53615.11}{\ac{MJD}}, some 1434 days before the glitch time $t_s^\xi$ (\SI{55048.92}{\ac{MJD}}).
The posterior distributions show that the modulation period and its first derivative experience fractional shifts of \SI{3.92E-04}{} and \SI{8.07E-01}{}, respectively.
To visualise this behaviour, in Fig.~\ref{fig:modulation_frequency_glitch}, we plot the inferred modulation period as a function of time.

To test the significance of our discovery that the modulation frequency, phase, and component amplitudes change at disjoint times spanning nearly the entire dataset, we repeat the analysis but restrict the prior distributions on the times of the step-changes to $+/-50$~days, centred on the reported glitch time.
We find that the Bayes factor between the full model and this restricted analysis is decisively in support of the full model with a natural-log Bayes factor of $877$.

\subsection{Model subsets}
\label{sec:other_models} 

To probe the relative importance of different features of the \textbf{S + P} model, we now explore two model subsets. First, a model which assumes that there is no glitch nor changes to the periodic modulation (\textbf{Model no-glitch}) and then a model which includes a step change only in the secular spin-down  (\textbf{Model S}). 

For \textbf{Model no-glitch}, we modify Eqn~\ref{eq:model_s_p} and \ref{eq:phase_glitch} removing the step changes leading to
\begin{equation}
    \dot{\nu}(t) = \sum_{i=0}^{N_s-1} \frac{\dot{\nu}_i}{i!}\Delta t^{i} + \sum_{j=1}^{N_c}A_j \cos\left(j\phi(t) + \Delta \phi_j\right) \,,
\end{equation}
and
\begin{equation}
    \phi(t) = 2\pi \sum_{k=0}^{N_f-1} \frac{1}{k !} f_k \Delta t^{k + 1} \,.
\end{equation}

Meanwhile, for \textbf{Model S}, which assumes a step-change at the glitch for the spin-down rate, we include a step change only in the secular part of the spin-down, i.e.:
\begin{align}
    \dot{\nu}(t) =  & \sum_{i=0}^{N_s-1} \frac{\dot{\nu}_i}{i!}\left[1 + H\left(t' -t_s^\xi\right)\left(\xi^p_i + \xi^t_i e^{-\frac{t' - t_s^\xi}{\tau_i}}\right)\right] \Delta t^{i} \nonumber \\ 
    & + \sum_{j=1}^{N_c}A_j \cos\left(j\phi(t) + \Delta \phi_j\right) \,,
\end{align}
with
\begin{equation}
    \phi(t) = 2\pi \sum_{k=0}^{N_f-1} \frac{1}{k !} f_k \Delta t^{k + 1} \,.
\end{equation}

\paragraph*{Model results}

The procedure described in Section~\ref{sec:glitch_model} was applied to both model subsets, from how the priors were defined to how the posterior distributions were obtained. Tables ~\ref{tab:priors_no_glitch_summary} and \ref{tab:priors_glitch_no_modulation_summary} list the full set of priors, for \textbf{Model no-glitch} and \textbf{Model S}, respectively. The choice of priors was the same as the ones presented in Section~\ref{sec:glitch_model} for \textbf{Model S+P}, but no glitch-related parameter priors and no modulation change-related parameter priors were included, for \textbf{Model no-glitch} and \textbf{Model S}, respectively.  

These subsets were obtained with $N_s = 4$, $N_c = 9$ and $N_f = 3$ and Tables ~\ref{tab:posterior_no_glitch_summary} and \ref{tab:posterior_glitch_no_modulation_summary} show that $f_2$, $A_9$ and $\dot{\nu}_3$ have maximum posterior probability values consistent with 0 within 1 $\sigma$, as was previously noted in \textbf{Model S+P}. \textbf{Model no-glitch} and \textbf{Model S} returned natural-log evidences of -- \SI{68308.4}{} $\pm$ 0.2 and \SI{68445.6}{} $\pm$ 0.2, respectively, lower than what was obtained for \textbf{Model S+P} (\SI{69931.9}{} $\pm$ 0.2). 

\begin{figure*}
    \centering
    \subfloat[]{
    \includegraphics[width=\linewidth]{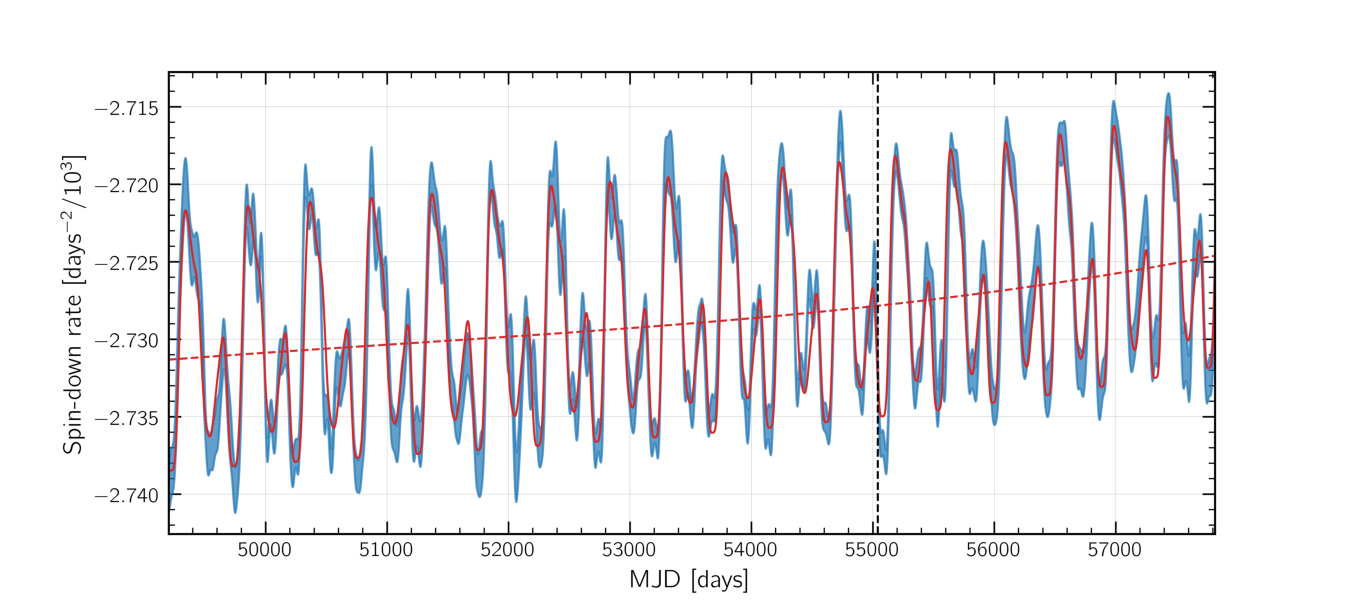}
    \label{fig:data_fit_no_glitch}} \\
    \subfloat[]{
     \includegraphics[width=\linewidth]{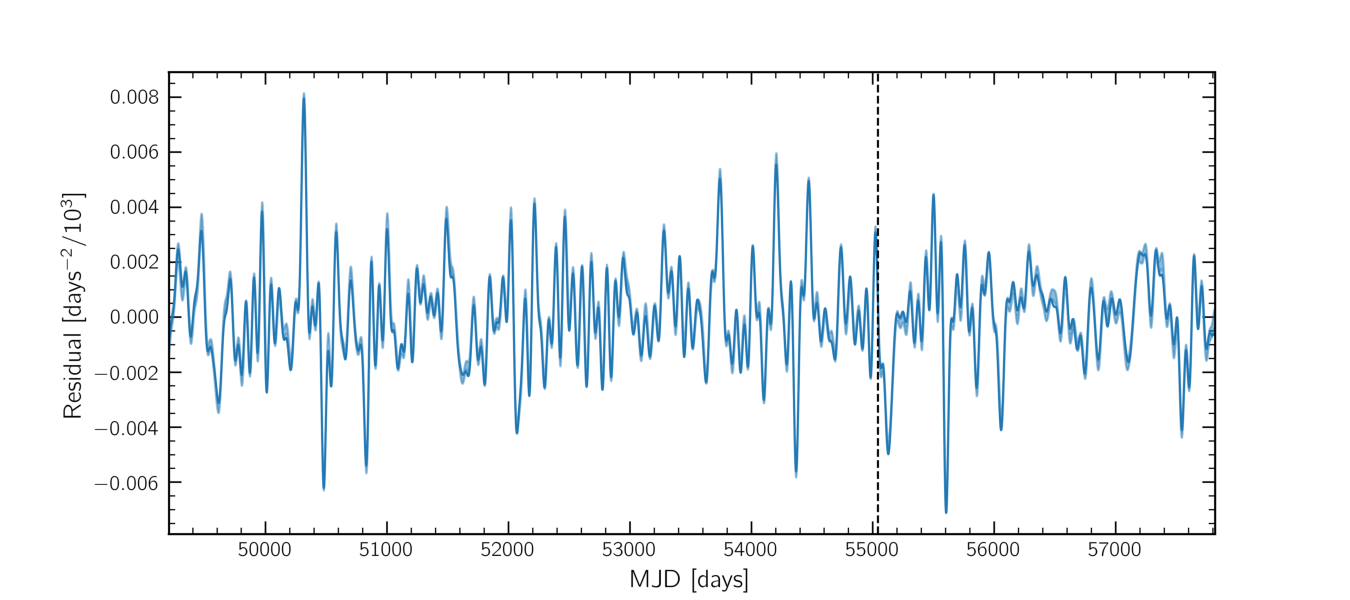}
    \label{fig:residuals_no_glitch}}
    \caption{\label{fig:data_fit_residuals_no_glitch} Figure similar to Fig.~\ref{fig:data_fit_residuals_glitch} showing the \ref{fig:data_fit_no_glitch} spin-down rate data together with the maximum posterior estimate solution of the model; and the \ref{fig:residuals_no_glitch} residuals, for \textbf{Model no-glitch}.}
\end{figure*}

\begin{figure*}
\centering
\subfloat[]{
\includegraphics[width=\linewidth]{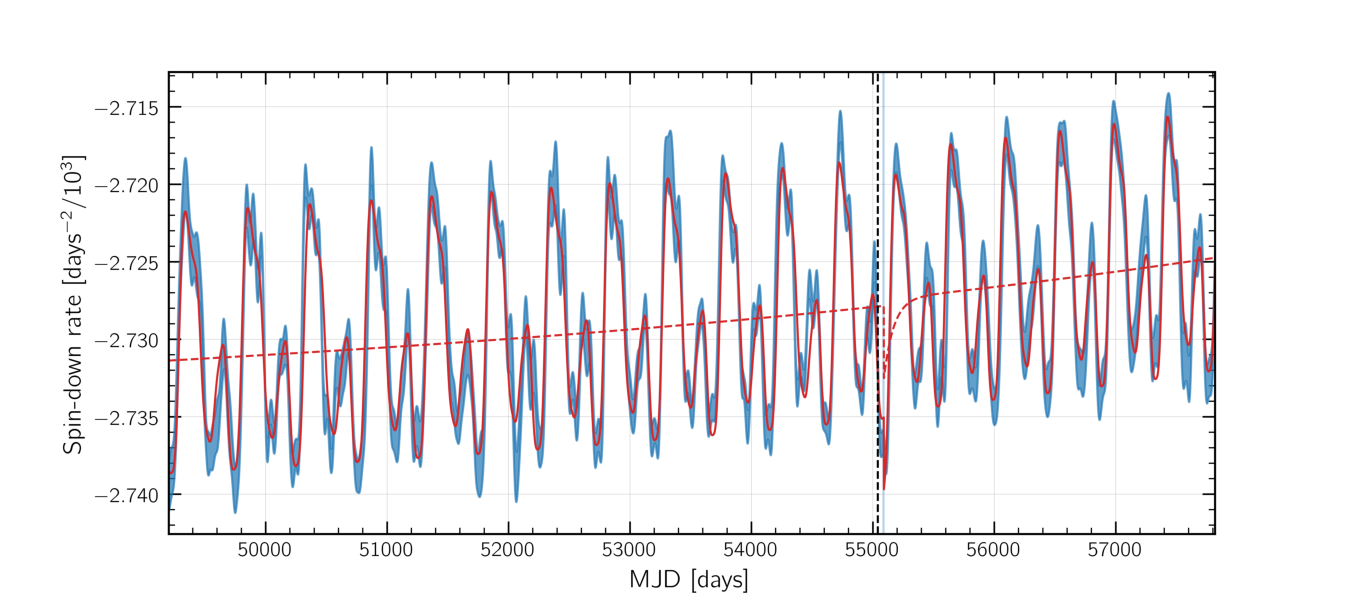}
\label{fig:data_fit_glitch_no_modulation}} \\
\subfloat[]{\includegraphics[width=\linewidth]{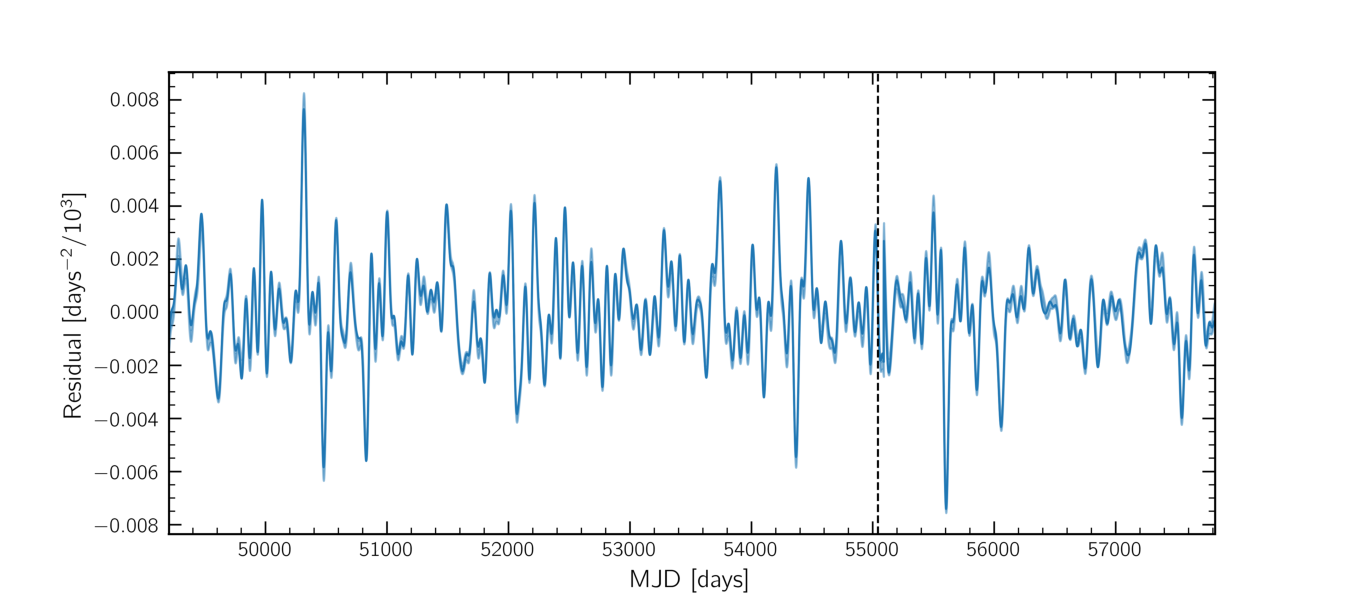}
\label{fig:residuals_glitch_no_modulation}}
\caption{\label{fig:data_fit_residuals_glitch_no_modulation} Figure similar to Fig.~\ref{fig:data_fit_residuals_glitch} showing the \ref{fig:data_fit_glitch_no_modulation} spin-down rate data together with the maximum posterior estimate solution of the model; and the \ref{fig:residuals_glitch_no_modulation} residuals, for \textbf{Model S}.}
\end{figure*}

Figures~\ref{fig:data_fit_no_glitch} and \ref{fig:data_fit_glitch_no_modulation} present the spin-down rate data (in blue) together with the maximum posterior estimate solution of the model (in red), which uses the parameters with the highest posterior probability, for \textbf{Model no-glitch} and \textbf{Model S}, respectively. These subset models were not able to capture the changes in the spin-down rate, in particular the transient recovery, that occurred after the glitch. This is evident in these figures but also in Figures~\ref{fig:residuals_no_glitch} and \ref{fig:residuals_glitch_no_modulation}, which show the residuals obtained by subtracting \textbf{Model no-glitch} and \textbf{Model S} from the data, respectively. By comparing these figures with Fig.~\ref{fig:residuals_glitch}, we can see that \textbf{Model no-glitch} and \textbf{Model S} are unable to capture the changes occurring on the data after the glitch, in the region between \SI{55000}{} and \SI{56000}{\ac{MJD}}.

\paragraph*{Discussion}
The model subsets perform poorly relative to \textbf{Model S+P} in modelling the observed spin-down rate of PSR B1828$-$11.
This is evidenced by the residual plots, which show larger deviations from zero and clear structures indicating specific instances where they fail, such as the transient recovery after the glitch, as addressed in the previous section. 
Moreover, we can perform a quantitative model comparison using the estimated natural-log evidence.
In Table~\ref{tab:model_bayes_factor}, we calculate the natural log-Bayes factors ($ln(K)$) demonstrating that \textbf{S+P} model is decisively preferred (e.g. using the interpretation from \citet{Kass1995}).
It is of note that the $ln(K)$ value obtained from \textbf{Models S+P} and \textbf{S} is lower than that obtained from \textbf{Models S+P} and \textbf{no-glitch}.
Since the models are nested, the S+P vs S Bayes factor can be compared to the S vs no-glitch Bayes factor to assess the relative importance of the secular glitch and the step changes in the periodic modulation.
Since the former is larger than the latter, this implies that for the spin-down data, the step changes in the modulation period are more significant than the secular changes.

\begin{table}
	\centering
	\caption{Tabulated $\ln$ Bayes factor, $ln(K)$, calculated for a comparison between \textbf{Model S+P} and the other models.}
	\label{tab:model_bayes_factor}
	\begin{tabular}{lcr} 
		\hline
		Model A & Model B & $\ln$(K) \\
		\hline
      \textbf{S+P} & \textbf{no-glitch} & 1623.60  \\
      \textbf{S+P} & \textbf{S} & 1486.34  \\
	\end{tabular}
\end{table}

\section{Comparing with model-independent visualisations}
\label{sec:lomb_scargle} 

In \citet{Ashton2017}, we introduced a time-period plot to study how the modulation period varies across the observed data span.
We now build on this concept in order to understand the implications of the \textbf{S+P} model inferences.
First, we fit and subtract a first-order polynomial from the raw spin-down rate data.
This ensures only the periodic modulations remain, and any information on the average spin-down rate or the second-order spin-down rate is removed.
We then plot the Lomb-Scargle periodogram \citep{Lomb1976,Scargle1982StudiesIA} applied in a sliding window with a stride length of \SI{1500}{days}. We varied this stride length, balancing long-duration windows that reduce the uncertainty on the estimation of the period with short-duration windows that increase the resolution in time.

\begin{figure*}
    \centering
    \includegraphics[width=\linewidth]{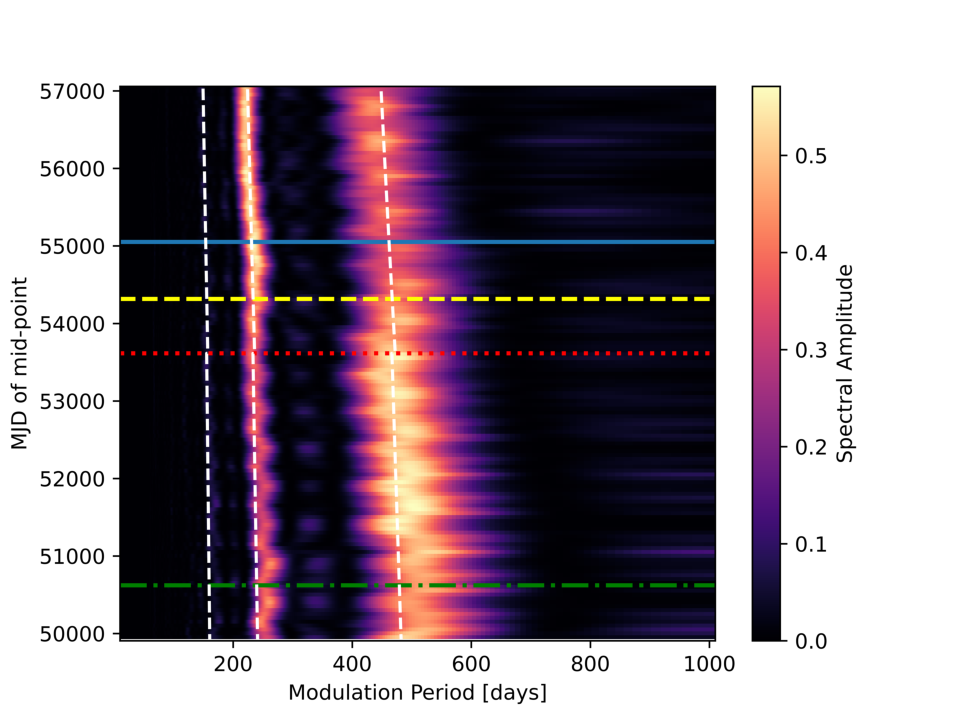}
    \caption{Modulation period spectrum of the spin-down rate residuals over a sliding window of \SI{1500}{days}, on the x-axis, as a function of the mid-point timestamp for each window, on the y-axis. The z-axis shows the Spectral Amplitude. The horizontal lines represent the glitch time parameters obtained by \textbf{Model S+P}.  $t_s^\xi$, in blue, represents the glitch time parameter; $t_s^\eta$, in yellow (dashed), represents the glitch time parameter related to a change in amplitude; $t_s^\chi$, in red (dotted), represents the glitch time parameter related to a change in modulation frequency and $t_s^\delta$, in green (dotted dashed), represents the glitch time parameter related to a change in phase offset. The vertical dashed white lines represent three modulation period modes returned by \textbf{Model S+P}.}
    \label{fig:lomb_scargle}
\end{figure*}

We find three modes in the spectrum: a primary mode at $\sim$ \SI{460}{days} (taking a reference epoch of \SI{50000}{\ac{MJD}}, the start of the data) and two smaller ones at $\sim$ \SI{250}{days} and $\sim$ \SI{170}{days}. The \SI{460}{days} and \SI{250}{days} modes have been modelled in \citet{Nitu2022} and \citet{Ashton2017}; the authors of \citet{Stairs2000} stated that there was a strong indication for the presence of a third mode at $\sim$ \SI{167}{days} and \citet{Rezania2003} confirmed the existence of this third harmonic. 

We add to Fig.~\ref{fig:lomb_scargle} horizontal lines denoting the epoch of the time parameters obtained by the data fit shown in Fig.~\ref{fig:data_fit_glitch}.
We also add white dashed lines to represent the three modulation periods and evolution of the three harmonic modes obtained by the fit on the spin-down rate data. They match the modulation periods obtained from the data. Higher derivatives of the modulation period obtained from the model are not displayed, as the Lomb-Scargle periodogram reveals no additional modes beyond those already presented.

From Fig.~\ref{fig:lomb_scargle}, we can clearly identify that the modulation period is decreasing over time and continues to do so after the glitch at approximately the same rate.
If we compare the modulation period value obtained for the major mode for the first and the last sliding window, we can see that the modulation period has decreased from \SI{489}{days} to \SI{435}{days}. From the Lomb-Scargle periodogram we can extract the maximum values and perform a linear regression across the entire dataset, as well as before and after the glitch. This calculation returned a rate of change of the modulation period of $\sim$ \SI{-0.011}{\second / \second}, consistent with the values previously calculated in \citet{Ashton2017} on the shorter pre-glitch data (see Fig.~\ref{fig:data_old_new_superimposed}). Additionally, we observe that before the glitch, the modulation period changes at a rate of $\sim$ \SI{-0.010}{\second / \second}. After the glitch, this rate increases to $\sim$ \SI{-0.014}{\second / \second}, indicating that the modulation period is decreasing more rapidly post-glitch. 

Another noticeable feature is the change of maximum spectral amplitude occurring at the same $t_s^\eta$ reported by the model, indicated by the line in yellow.
A decrease in the spectral amplitude occurs at $t_s^\eta$ for the first modulation period mode, while the second modulation period mode sees an increase.
This is also consistent with the inferences made from \textbf{Model S+P}.
Finally, a shift in the modulation period and frequency is observed at $t_s^\chi$, matching the model, indicated by the  line in red. After this point, the modulation period temporarily shifts to the right, indicating an increase. Subsequently, it resumes its continuous decrease for the remainder of the data range. This behaviour aligns with the model results depicted in Fig.~\ref{fig:modulation_frequency_glitch}.

\section{Discussion}
\label{sec:discussion}

In this work, we analyse a newly available high-resolution and expanded data set containing the spin-down rate of PSR B1828$-$11.
The longer data set contains several cycles of observations after the pulsar glitch at \SI{55040.9}{\ac{MJD}}.
As opposed to previous efforts, in which physics-informed models were developed to explain the behaviour of the pulsar, here we apply a phenomenological model. We considered three models to describe the behaviour of the pulsar: \textbf{Model S+P}, which considered the existence of a glitch and changes to the periodic modulation of the star; \textbf{Model no-glitch}, which assumed that no glitch nor changes to the periodic modulation occurred; and \textbf{Model S}, which allowed for a glitch but assumed no modulation changes. \textbf{Model S+P} was preferred over the other two, as detailed in Section~\ref{sec:other_models}, with the Bayes factor values shown in Table~\ref{tab:model_bayes_factor}. 

\textbf{Model S+P} was designed to allow for parameter changes at the glitch, with a preference for no change to occur, by using ‘Slab-and-spike’ priors.  For each component of the model, i.e. the spin-down rate, the amplitude, the phase-offset and the modulation frequency, we allowed for a separate parameter recording the time at which the step occurred, to see if these changes coincided with the glitch time. This was not the case, with step changes for some parameters being found to occur before the glitch: a decrease of modulation amplitude, described by $\eta_1$ = -0.175 $\pm$ 0.008 was 
observed for the main modulation period mode at $t_s^\eta$; a change in modulation frequency was observed, as seen in Fig.~\ref{fig:modulation_frequency_glitch}, at $t_s^\chi$; all cosine components observed a change in the phase-offset at $t_s^\delta$. Additionally, we found that the spin-down rate experienced an exponential recovery after the glitch $t_s^\xi$.

To add validity to this model, the data was analysed without a model, as described in Section~\ref{sec:lomb_scargle}. We obtained a Lomb Scargle periodogram spectrum from the data. This shows that the modulation period of the pulsar is decreasing at a rate of $\sim$ \SI{-0.010}{\second / \second} before the glitch and decreasing more rapidly after the glitch ($\sim$ \SI{-0.014}{\second / \second}); the modulation period modes experienced spectral amplitude changes before the glitch time; and we visually identify a step change to the modulation frequency and thus the modulation period, also before the glitch. 

\subsection{Interpreting the results in light of astrophysical models}
\paragraph*{Planetary companions:} It is clear already from the correlated changes in the spin-down and pulse shape that a planetary explanation for the modulations is unlikely.
Nevertheless, the observation of a decreasing modulation period \citep{Ashton2017} invites reconsideration of the planetary explanation, with some coupling torque between the star and planet(s) yielding the inspiral and explaining the changing modulation period.
However, while no generative model exists, we feel that any such model would necessarily require the modulations to arise from the smoothly varying orbital separation of the planets and star.
Therefore, the observation in this work is that there are distinct instantaneous changes in the modulation seems difficult to explain with a planetary hypothesis.
Moreover, the observation that there are up to 8 harmonically related sinusoids would also call for up to 8 planets, following the original arguments that the two sinusoids correspond to two planets. Once again, this feels implausible.

\paragraph*{Free precession}

The global difficulties of reconciling free precession, the decreasing modulation period, and the glitch have already been discussed in \citet{Jones2017}.
Here, we can quantify the observation from \citet{Shaw2022, Stairs2019} that contrary to some of the predictions of \citet{Jones2017}, the modulations continue after the glitch, constraining the models connecting the interior of the star to the cause of the modulations.
Furthermore, the observation that there are distinct changes (at different times) in the modulation amplitude, frequency, and phase adds to the challenge of interpreting this event in light of free precession.
However, the observation of multiple harmonically related sinusoids offers a new opportunity to test the model.
Namely, in \citet{Jones2001}, the precession model used in later works was developed with an expansion in the small angle $\theta$ between the symmetry axis of the (assumed biaxial) moment of inertia tensor and the angular momentum.
Therefore, a more physically accurate model can be obtained by either expanding the model to include higher-order terms.
Since the precession model has no additional degrees of freedom left, it will be interesting to discover if the amplitude coefficients of the harmonically related sinusoids measured here are consistent with the predictions of precession, enabling a new test of precession.
However, to explain the observed beam-width data, such a model would need to be extended as suggested by \citet{Stairs2019} to combine the long-term precession behavior with quantized profile switches.

\paragraph*{Magnetospheric switching}
There is no single well-defined magnetospheric switching model: in a sense, it is a set of observations rather than a generative model itself.
It is, therefore, not straightforward to connect our observations to such a model.
Moreover, since we are using only the spin-down rate data, we are insensitive to the rapid switches between states and can only discuss the long-timescale periodic modulations in this switching rate. 
Nevertheless, our phenomenological study reveals several insights into whatever process drives this. First, there is decisive evidence for more than two terms in the harmonic expansion; this is observable directly by our fit to the data but also by the non-zero posterior support for these terms. The amplitude of these terms could provide a way to test mechanisms for the clock (e.g. as proposed for precession in the previous section). 
Second, we find evidence for distinct changes in the spectral amplitude of the harmonic sinusoid, with a sudden shift from the fundamental to the first harmonic.
This is interesting as it suggests another variability mechanism for the periodic modulations. It would be interesting to study the raw data of PSR B1828$-$11 to identify if there are corresponding systematic changes in the beam shape during this transition (or, indeed, any of the observed step changes). 

In \citet{Seymour2013}, the authors introduced evidence that PSR B1828$-$11 was displaying chaotic behaviour consistent with a system with 3 governing variables.
From this work and further discussion \citet{Stairs2019}, it is proposed that the spin-down rate and mode transition rate act as two of the governing variables.
However, it is unclear see how this observation can be connected to a physical process to predict the observed chaotic behaviour.

\section{Outlook}
\label{sec:outlook}
The high-quality data released by \citet{Nitu2022} has enabled a new and detailed study of PSR B1828$-$11 using the inferred spin-down rate.
Since the end of the data set studied in this work, \ac{JBO} has continued observing PSR B1828$-$11 and we expect there to be several more cycles to study. 
Moreover, there are other pulsars which display similar (if less clear) behaviours. For example, the authors of \citet{zubieta2024glitchinducedpulseprofilechange} have reported on changes to the amplitude and frequency of PSR J0742$-$2822 following a glitch.
We believe the tools and techniques developed in this work could be applied to larger data sets, with the ultimate goal of providing quantitative measures of the behaviour to help us constrain models.
However, one key missing aspect is that we are studying only the spin-down rate and neglecting information about the mode-transition rate.
Therefore, we believe further methodological work is needed to develop approaches that can automate the analysis of pulsars. This would allow the study of both their rapidly changing beam shape and their long-term timing properties. We believe this has the capacity to answer long-held questions about the star's interaction with its magnetosphere.

The work presented here shows that a model that allows for sudden step changes in the amplitude, phase and frequency parameters in the long-term periodicity in the spin-down of PSR B1828$-$11 is a better fit to the data than a model that does not allow for sudden step changes  in these parameters.  Significantly, the model locates these step changes at \emph{three different times}, all well \emph{before} the glitch itself.  This is somewhat surprising, and difficult to account for in terms of a physical model.  This suggests that other models of the long-term  periodicity, not based on sudden step changes, may also be worth exploring.

\section*{Acknowledgements}

We utilise the \texttt{Numpy} \citep{harris2020array} and \texttt{Matplotlib} \citep{Hunter:2007} libraries for data processing and visualisation and the \texttt{Scipy} library \citep{2020SciPy-NMeth} for implementation of the Lomb-Scargle periodogram.  DĲ acknowledges support from the Science and Technology Funding Council (STFC)  via grant No. ST/R00045X/1.

\section*{Data Availability}

The data used in this publication is available in \citet{keith_2023_7664166}.     
The code needed to produce the results shown in this publication can be shared upon request to the corresponding author.



\bibliographystyle{mnras}
\bibliography{bibliography} % if your bibtex file is called example.bib

%%%%%%%%%%%%%%%%% APPENDICES %%%%%%%%%%%%%%%%%%%%%

\appendix

\section{Prior distributions and summary statistics. }

\begin{table}
	\centering
	\caption{Prior distributions for the \textbf{Model S+P}'s parameters. Parameters with priors denoted with 'SS' have slab-spike priors applied to them}
	\label{tab:priors_glitch_summary}
	\begin{tabular}{lcr} % four columns, alignment for each
		\hline
		 & Prior & Units\\
		\hline
$\dot{\nu}_0$	&	Unif(	-2.74$\times10^{-3}$	,	-2.72$\times10^{-3}$	)	&	$\SI{}{days^{-2}}$	\\
$\dot{\nu}_1$	&	Unif(	-2.73$\times10^{-7}$	,	2.73$\times10^{-7}$	)	&	$\SI{}{days^{-3}}$	\\
$\dot{\nu}_2$	&	Unif(	-2.73$\times10^{-11}$	,	2.73$\times10^{-11}$	)	&	$\SI{}{days^{-4}}$	\\
$\dot{\nu}_3$	&	Unif(	-2.73$\times10^{-15}$	,	2.73$\times10^{-15}$	)	&	$\SI{}{days^{-5}}$	\\
$\tau$	&	Unif(	0	,	500	)	&	$\SI{}{days}$	\\
$t_s^\xi$	&	Unif(	54990.90	,	55090.90	)	&	$\SI{}{days}$	\\
$\xi_p^0$ 	&	SS + Unif(	-0.01	,	0.01	)	&	$\SI{}{days}$	\\
$\xi_t^0$ 	&	SS + Unif(	-0.01	,	0.01	)	&	$\SI{}{days}$	\\
$A_{1-9}$ 	&	SS + Unif(	0	,	1.00$\times10^{-5}$	)	&	-	\\
$\eta_{1-9}$ 	&	SS + $\mathcal{N}$(	0	,	1	)	&	-	\\
$\Delta\phi_{1-9}$	&	Unif(	$-\pi$	,	$\pi$	)	&	$\SI{}{\radian}$	\\
$\delta_{1-9}$ 	&	SS + $\mathcal{N}$(	0	,	1	)	&	-	\\
$f_0$ 	&	SS + Unif(	2.11$\times10^{-3}$	,	2.33$\times10^{-3}$	)	&	$\SI{}{days^{-1}}$	\\
$f_1$ 	&	SS + Unif(	-2.22$\times10^{-7}$	,	2.22$\times10^{-7}$	)	&	$\SI{}{days^{-2}}$	\\
$f_2$ 	&	SS + Unif(	-2.22$\times10^{-11}$	,	2.22$\times10^{-11}$	)	&	$\SI{}{days^{-3}}$	\\
$\chi_0$ 	&	SS + Unif(	-1	,	1	)	&	-	\\
$\chi_1$ 	&	SS + Unif(	-1	,	1	)	&	-	\\
$\chi_2$ 	&	SS + Unif(	-1	,	1	)	&	-	\\
$\Delta t_s^\chi$ 	&	SS + Unif(	-5000	,	2000	)	&	-	\\
$\Delta t_s^\delta$ 	&	SS + Unif(	-5000	,	2000	)	&	-	\\
$\Delta t_s^\eta$ 	&	SS + Unif(	-5000	,	2000	)	&	-	\\
	\end{tabular}
\end{table}

\begin{table}
	\centering
	\caption{Prior distributions for \textbf{Model no-glitch}'s  parameters. Parameters with priors denoted with 'SS' have slab-spike priors applied to them}
	\label{tab:priors_no_glitch_summary}
	\begin{tabular}{lcr} % four columns, alignment for each
		\hline
		 & Prior & Units\\
		\hline
$\dot{\nu}_0$	&	Unif(	-2.74$\times10^{-3}$	,	-2.72$\times10^{-3}$	)	&	$\SI{}{days^{-2}}$	\\
$\dot{\nu}_1$	&	Unif(	-2.73$\times10^{-7}$	,	2.73$\times10^{-7}$	)	&	$\SI{}{days^{-3}}$	\\
$\dot{\nu}_2$	&	Unif(	-2.73$\times10^{-11}$	,	2.73$\times10^{-8}$	)	&	$\SI{}{days^{-4}}$	\\
$\dot{\nu}_3$	&	Unif(	-2.73$\times10^{-15}$	,	2.73$\times10^{-8}$	)	&	$\SI{}{days^{-5}}$	\\
$A_{1-9}$ 	&	SS + Unif(	0	,	1.00$\times10^{-5}$	)	&	-	\\
$\Delta\phi_{1-9}$	&	Unif(	$-\pi$	,	$\pi$	)	&	$\SI{}{\radian}$	\\
$f_0$	&	SS + Unif(	2.11$\times10^{-3}$	,	2.33$\times10^{-3}$	)	&	$\SI{}{days^{-1}}$	\\
$f_1$	&	SS + Unif(	-2.22$\times10^{-7}$	,	2.22$\times10^{-7}$	)	&	$\SI{}{days^{-2}}$	\\
$f_2$	&	SS + Unif(	-2.22$\times10^{-11}$	,	2.22$\times10^{-11}$	)	&	$\SI{}{days^{-3}}$	\\
	\end{tabular}
\end{table}

\begin{table}
	\centering
	\caption{Prior distributions for \textbf{Model S}'s parameters. Parameters with priors denoted with 'SS + Unif' have slab-spike priors applied to them}
	\label{tab:priors_glitch_no_modulation_summary}
	\begin{tabular}{lccr} % four columns, alignment for each
		\hline
		& Prior & Units\\
		\hline
      $\dot{\nu}_0$	&	Unif(	-2.74$\times10^{-3}$	,	-2.72$\times10^{-3}$	)	&	$\SI{}{days^{-2}}$	\\
$\dot{\nu}_1$	&	Unif(	-2.73$\times10^{-7}$	,	2.73$\times10^{-8}$	)	&	$\SI{}{days^{-3}}$	\\
$\dot{\nu}_2$	&	Unif(	-2.73$\times10^{-11}$	,	2.73$\times10^{-11}$	)	&	$\SI{}{days^{-4}}$	\\
$\dot{\nu}_3$	&	Unif(	-2.73$\times10^{-15}$	,	2.73$\times10^{-15}$	)	&	$\SI{}{days^{-5}}$	\\
$\tau$	&	Unif(	0	,	500	)	&	$\SI{}{days}$	\\
$t_s^\xi$	&	Unif(	54990.90	,	55090.90	)	&	$\SI{}{days}$	\\
$\xi_p^0$	&	SS + Unif(	-0.01	,	0.01	)	&	$\SI{}{days}$	\\
$\xi_t^0$	&	SS + Unif(	-0.01	,	0.01	)	&	$\SI{}{days}$	\\
$A_{1-9}$	&	SS + Unif(	0	,	1.00$\times10^{-5}$	)	&	-	\\
$\Delta\phi_{1-9}$	&	Unif(	$-\pi$	,	$\pi$	)	&	$\SI{}{\radian}$	\\
$f_0$	&	SS + Unif(	2.11$\times10^{-3}$	,	2.33$\times10^{-3}$	)	&	$\SI{}{days^{-1}}$	\\
$f_1$	&	SS + Unif(	-2.22$\times10^{-7}$	,	2.22$\times10^{-7}$	)	&	$\SI{}{days^{-2}}$	\\
$f_2$	&	SS + Unif(	-2.22$\times10^{-11}$	,	2.22$\times10^{-11}$	)	&	$\SI{}{days^{-3}}$	\\

	\end{tabular}
\end{table}

\begin{table}
	\centering
	\caption{Maximum posterior distribution summary, with their standard deviations, for \textbf{Model S+P}'s parameters}
	\label{tab:posterior_glitch_summary}
	\begin{tabular}{lccr} % four columns, alignment for each
		\hline
		 & Posterior median(s.d.) & Units\\
		\hline
$\dot{\nu}_0$	&	-2.72739(5)$\times10^{-3}$	&	$\SI{}{days^{-2}}$	\\
$\dot{\nu}_1$	&	9.0(3)$\times10^{-10}$	&	$\SI{}{days^{-3}}$	\\
$\dot{\nu}_2$	&	6(2)$\times10^{-14}$	&	$\SI{}{days^{-4}}$	\\
$\dot{\nu}_3$	&	-2(1)$\times10^{-17}$	&	$\SI{}{days^{-5}}$	\\
$\tau$	&	121(14)	&	$\SI{}{days}$	\\
$t_s^\xi$	&	55047(3) &	$\SI{}{days}$	\\
$\xi_p^0$	&	-5(3)$\times10^{-5}$	&	$\SI{}{days}$	\\
$\xi_t^0$	&	1.23(8)$\times10^{-3}$	&	$\SI{}{days}$	\\
$A_1$	&	5.77(3)$\times10^{-6}$	&	-	\\
$A_2$	&	4.77(3)$\times10^{-6}$	&	-	\\
$A_3$	&	2.03(3)$\times10^{-6}$	&	-	\\
$A_4$	&	9.3(3)$\times10^{-7}$	&	-	\\
$A_5$	&	7.4(3)$\times10^{-7}$	&	-	\\
$A_6$	&	2.0(3)$\times10^{-7}$	&	-	\\
$A_7$	&	5.8(3)$\times10^{-7}$	&	-	\\
$A_8$	&	2.6(3)$\times10^{-7}$	&	-	\\
$A_9$	&	0(1)$\times10^{-9}$	&	-	\\
$\eta_1$	&	-0.175(8)	&	-	\\
$\eta_2$	&	0.19(1)	    &	-	\\
$\eta_3$	&	-0.38(2)	&	-	\\
$\eta_4$	&	-0.32(5)    &	-	\\
$\eta_5$	&	0(3)	    &	-	\\
$\eta_6$	&	0.8(4)	    &	-	\\
$\eta_7$	&	-0.30(9)	&	-	\\
$\eta_8$	&	-0.8(1) 	&	-	\\
$\eta_9$	&	-0.03(7)	&	-	\\
$\Delta\phi_1$	&	2.30(1)	    &	$\SI{}{\radian}$	\\
$\Delta\phi_2$	&	-1.48(2)	&	$\SI{}{\radian}$	\\
$\Delta\phi_3$	&	3.139(9)	&	$\SI{}{\radian}$	\\
$\Delta\phi_4$	&	-0.32(3)	&	$\SI{}{\radian}$	\\
$\Delta\phi_5$	&	1.4(1)	    &	$\SI{}{\radian}$	\\
$\Delta\phi_6$	&	-0.6(2) 	&	$\SI{}{\radian}$	\\
$\Delta\phi_7$	&	3.13(4)	    &	$\SI{}{\radian}$	\\
$\Delta\phi_8$	&	2.7(1)	    &	$\SI{}{\radian}$	\\
$\Delta\phi_9$	&	2(2)	    &	$\SI{}{\radian}$	\\
$\delta_1$	&	-0.143(5)	    &	-	\\
$\delta_2$	&	0.39(1)	        &	-	\\
$\delta_3$	&	-0.276(6)	    &	-	\\
$\delta_4$	&	4.4(4)	        &	-	\\
$\delta_5$	&	-0.59(5)    	&	-	\\
$\delta_6$	&	3.7(5)      	&	-	\\
$\delta_7$	&	-0.47(2)    	&	-	\\
$\delta_8$	&	0(4)	        &	-	\\
$\delta_9$	&	1.5(7)	        &	-	\\
$f_0$	&	2.1748(3)$\times10^{-3}$	&	$\SI{}{days^{-1}}$	\\
$f_1$	&	1.847(7)$\times10^{-8}$	    &	$\SI{}{days^{-2}}$	\\
$f_2$	&	0(6)$\times10^{-16}$	    &	$\SI{}{days^{-3}}$	\\
$\chi_0$	&	3.9(2)$\times10^{-4}$	&	-	\\
$\chi_1$	&	0.81(2)	                &	-	\\
$\chi_2$	&	0.2(4)              	&	-	\\
$\Delta t_s^\chi$	&	-1424(4)	&	-	\\
$\Delta t_s^\delta$	&	-4420(20)	&	-	\\
$\Delta t_s^\eta$	&	-740(6)	&	-	\\
	\end{tabular}
\end{table}

\begin{table}
	\centering
	\caption{Maximum posterior distribution summary, with their standard deviations, for \textbf{Model no-glitch}'s parameters}
	\label{tab:posterior_no_glitch_summary}
	\begin{tabular}{lccr} % four columns, alignment for each
		\hline
		 & Posterior median(s.d.) & Units\\
		\hline     
$\dot{\nu}_0$	&	-2.72754(4)$\times10^{-3}$	&	$\SI{}{days^{-2}}$	\\
$\dot{\nu}_1$	&	9.5(2)$\times10^{-10}$	&	$\SI{}{days^{-3}}$	\\
$\dot{\nu}_2$	&	1.8(2)$\times10^{-13}$	&	$\SI{}{days^{-4}}$	\\
$\dot{\nu}_3$	&	3(1)$\times10^{-17}$	&	$\SI{}{days^{-5}}$	\\
$A_1$	&	5.27(3)$\times10^{-6}$ &	-	\\
$A_2$	&	5.08(3)$\times10^{-6}$	&	-	\\
$A_3$	&	1.66(3)$\times10^{-6}$	&	-	\\
$A_4$	&	7.1(3)$\times10^{-7}$	&	-	\\
$A_5$	&	6.6(3)$\times10^{-7}$	&	-	\\
$A_6$	&	2.5(3)$\times10^{-7}$	&	-	\\
$A_7$	&	3.7(3)$\times10^{-7}$	&	-	\\
$A_8$	&	9(5)$\times10^{-8}$	    &	-	\\
$A_9$	&	0(3)$\times10^{-9}$	    &	-	\\
$\Delta\phi_1$	&	2.070(7)	&	$\SI{}{\radian}$	\\
$\Delta\phi_2$	&	-1.875(8)	&	$\SI{}{\radian}$	\\
$\Delta\phi_3$	&	2.53(2)	    &	$\SI{}{\radian}$	\\
$\Delta\phi_4$	&	-1.16(5)	&	$\SI{}{\radian}$	\\
$\Delta\phi_5$	&	0.85(5) 	&	$\SI{}{\radian}$	\\
$\Delta\phi_6$	&	-1.6(1) 	&	$\SI{}{\radian}$	\\
$\Delta\phi_7$	&	2.6(1)	    &	$\SI{}{\radian}$	\\
$\Delta\phi_8$	&	3(2)	    &	$\SI{}{\radian}$	\\
$\Delta\phi_9$	&	2(2)    	&	$\SI{}{\radian}$	\\
$f_0$	&	2.1874(3)$\times10^{-3}$	&	$\SI{}{days^{-1}}$	\\
$f_1$	&	2.16(1)$\times10^{-8}$	&	$\SI{}{days^{-2}}$	\\
$f_2$	&	0(1)$\times10^{-15}$	&	$\SI{}{days^{-3}}$	\\

	\end{tabular}
\end{table}

\begin{table}
	\centering
	\caption{Maximum posterior distribution summary, with their standard deviations, for \textbf{Model S}'s parameters}
	\label{tab:posterior_glitch_no_modulation_summary}
	\begin{tabular}{lccr} % four columns, alignment for each
		\hline
		& Posterior median(s.d.) & Units\\
		\hline
$\dot{\nu}_0$	&	-2.72737(7)$\times10^{-3}$	&	$\SI{}{days^{-2}}$	\\
$\dot{\nu}_1$	&	9.2(4)$\times10^{-10}$	&	$\SI{}{days^{-3}}$	\\
$\dot{\nu}_2$	&	9(2)$\times10^{-14}$	&	$\SI{}{days^{-4}}$	\\
$\dot{\nu}_3$	&	-0.03(1)$\times10^{-17}$	&	$\SI{}{days^{-5}}$	\\
$\tau$	&	74(8)	&	$\SI{}{days}$	\\
$t_s^\xi$	&	55091(2)	&	$\SI{}{days}$	\\
$\xi_p^0$	&	-3.67(4)$\times10^{-5}$	&	$\SI{}{days}$	\\
$\xi_t^0$	&	2.1037(1)$\times10^{-3}$	&	$\SI{}{days}$	\\
$A_1$	&	5.32(3)$\times10^{-6}$	&	-	\\
$A_2$	&	5.07(3)$\times10^{-6}$	&	-	\\
$A_3$	&	1.67(3)$\times10^{-6}$	&	-	\\
$A_4$	&	6.8(3)$\times10^{-7}$	&	-	\\
$A_5$	&	6.7(3)$\times10^{-7}$	&	-	\\
$A_6$	&	2.3(3)$\times10^{-7}$	&	-	\\
$A_7$	&	3.6(3)$\times10^{-7}$	&	-	\\
$A_8$	&	1.4(6)$\times10^{-7}$	&	-	\\
$A_9$	&	0(4)$\times10^{-9}$	&	-	\\
$\Delta\phi_1$	&	2.088(6)	&	$\SI{}{\radian}$	\\
$\Delta\phi_2$	&	-1.853(7)	&	$\SI{}{\radian}$	\\
$\Delta\phi_3$	&	2.57(2)  	&	$\SI{}{\radian}$	\\
$\Delta\phi_4$	&	-1.11(5)	&	$\SI{}{\radian}$	\\
$\Delta\phi_5$	&	0.88(5)	    &	$\SI{}{\radian}$	\\
$\Delta\phi_6$	&	-1.7(1)	    &	$\SI{}{\radian}$	\\
$\Delta\phi_7$	&	2.52(9)	    &	$\SI{}{\radian}$	\\
$\Delta\phi_8$	&	3(2)	    &	$\SI{}{\radian}$	\\
$\Delta\phi_9$	&	2(2)	    &	$\SI{}{\radian}$	\\
$f_0$	&	2.1868(3)$\times10^{-3}$	&	$\SI{}{days^{-1}}$	\\
$f_1$	&	2.14(1)$\times10^{-8}$   	&	$\SI{}{days^{-2}}$	\\
$f_2$	&	0(1)$\times10^{-16}$ 	&	$\SI{}{days^{-3}}$	\\

	\end{tabular}
\end{table}

%%%%%%%%%%%%%%%%%%%%%%%%%%%%%%%%%%%%%%%%%%%%%%%%%%

% Don't change these lines
\bsp	% typesetting comment
\label{lastpage}
\end{document}